\documentclass[aps,prl,floatfix,twocolumn,showpacs]{revtex4}
\usepackage{graphicx} 
\usepackage[sort&compress]{natbib}
\usepackage{graphicx}
\newcommand{\BEQ}{\begin{equation}}
\newcommand{\EEQ}{\end{equation}}
\newcommand{\BEA}{\begin{eqnarray}}
\newcommand{\EEA}{\end{eqnarray}}
\newcommand{\nn}{\nonumber \\}
\renewcommand{\k}{\alpha}
\renewcommand{\d}{{\rm d}}

\renewcommand{\nn}{\nonumber}

\newcommand{\half}{\frac{1}{2}}

\newcommand{\wphi}{\zeta}

\newcommand{\pdk}{\vec{\partial}_{\k}}

\newcommand{\ka}{\kappa}
\begin{document}
\draft
\title
{Adhesion Induced DNA Naturation}
\date{\today}

\author{A.E. Allahverdyan$^{1,2)}$, Zh.S. Gevorkian$^{1,3,4)}$,
Chin-Kun Hu$^{4)}$ and Th.M. Nieuwenhuizen$^{2)}$
}

\address{$^{1)}$Yerevan Physics Institute,
Alikhanian Brothers St. 2, Yerevan 375036, Armenia,\\
$^{2)}$ Institute for Theoretical Physics, 
Valckenierstraat 65, 1018 XE Amsterdam, The Netherlands,\\
$^{3)}$ Institute of Radiophysics and
Electronics, Ashtarak-2, 378410, Armenia,\\
$^{4)}$ Institute of Physics, Academia of Sinica, Nankang, Taipei
11529, Taiwan.
}

\begin{abstract} DNA adsorption and naturation is modeled via two
interacting flexible homopolymers coupled to a solid surface.
DNA denatures if the entropy gain for unbinding the two strands
overcomes the loss of binding energy.  When adsorbed to a surface, the
entropy gain is smaller than in the bulk, leading to a stronger binding
and, upon neglecting self-avoidance, absence of a denatured phase.  Now
consider conditions where the binding potentials are too weak for
naturation, and the surface potential too weak to adsorb single strands.
In a variational approach it is shown that their combined action may
lead to a naturated adsorbed phase.  Conditions for the absence of
naturation and adsorption are derived too.  The phase diagram is
constructed qualitatively.

\end{abstract}


\maketitle

The structure of biopolymers is the key for understanding their
biological functioning \cite{bio,GKH}.  A known component of this
structure is the double-strand|two polymers lined up by hydrogen
bonds|realized in DNA and responsible for stability of the genetic information.

We shall study the adsorption and surface (adhesion) induced
naturation of a double-strand polymer with resolved motion of the
separate strands.  The adsorption of DNA onto surfaces is of interest
for a number of applications such as design of substrates as carriers of
genetically engineered vaccines, surface patterning, histone-DNA interactions, and
electrophoresis \cite{review}.  DNA adsorption can be realized by different mechanisms
\cite{review}: covalent bonding, electrostatic attraction to a positively
charged surface \cite{SPM}, hydrogen bonding, and
hydrophobic forces.

When the two strands are tightly connected to each other, the
double-strand is described via a single semi-flexible chain
\cite{deG,singlestiff,lip}. Indeed, for the double-stranded DNA the
corresponding persistence length $l_p$ is about $100$ nm \cite{stiff},
or $300$ base-pairs.  However, this large value of $l_p$ is more a
result of ``construction rigidity'', because for a single DNA strand
$l_p$ is much smaller, about 1 nm (2-3 base-pairs) \cite{free}, while
the length of the strand is 10-100 $\mu$m.  The single DNA strand can
thus be modeled as a flexible chain \cite{deG}.  Adsorption of a
(semi)flexible chain is by now a classical problem in polymer physics
\cite{GKH,deG,singlestiff}.

If the motion of the single strands is resolved, i.e., when the
inter-strand hydrogen bonds are relatively weak, the polymer becomes
a complex system with different, mutually balancing features.
A realistic model of DNA should include
stacking energy between two base pairs, the helical structure,
self-avoidance, heterogeneity, etc.  Such models do not seem to
exist; there are, however, various models with different degrees of
sophistication capturing at least some aspects
of the problem \cite{GKH,deG,singlestiff,lip,PB,PolScher,Grass}.

Our model deals only with most basic features of
the problem.  We consider two flexible Gaussian chains interacting with
each other and with a plane solid surface. 
Though neglecting some features of DNA (see below), our 
model predicts two mechanisms of adhesion induced naturation.
This model without
surface was mentioned in \cite{deG} and studied in \cite{PB}
in the context of DNA denaturation (melting); see also 
\cite{bio,GKH,PolScher,Grass} for recent results.


Consider $2N$ coupled monomers with
radius-vectors $\vec{r}_{1| k}$ and $\vec{r}_{2| k}$
($k=1,...,N$) and potential energy
\begin{eqnarray}
\sum_{k=1}^N\left\{{\cal U}(\vec r_{k})+
\sum_{\alpha=1}^2
\left[\frac{l}{2}
\left(\vec{r}_{\alpha| k}-\vec{r}_{\alpha| k-1}\right)^2
+{\cal V}(\vec{r}_{\alpha| k})
\right]\right\},~\nonumber
\label{Ham}
\end{eqnarray}
where $\vec r_{k}\equiv \vec r_{1| k}-\vec r_{2| k}$, ${\cal U}$ is the
inter-strand potential and ${\cal V}$ is the surface-monomer potential.
The harmonic interaction with stiffness $l$ is responsible for the
linear structure of the polymers.  The system is embedded in an
equilibrium thermal bath at temperature $T=1/\beta$ ($k_{\rm B}=1$).
The quadratic kinetic energy is irrelevant, since it factorizes from the
partition function and does not influence the monomer density.  It is
known that in the thermodynamical limit $N\gg 1$ the free energy of
a single flexible chain (without self-interactions) is determined from
an effective Schr\"odinger equation \cite{GKH,deG}.  The considered
two-strand situation is a direct generalization of this, with the
two-particle equation being $H\Psi=E\Psi$, 
\BEA
\label{1} H\equiv{\sum}_{\k=1}^2[
-\half\,\pdk^{\,2}
+ V(z_\k)]+U(r),\quad
\pdk\equiv\partial_{\vec{r}_{\k}}.
  \EEA
Here $V\equiv l\beta^2{\cal V}$, $U\equiv l\beta^2{\cal U}$,
while $\vec{r}_{1,2}=(x_{1,2},y_{1,2},z_{1,2})$
are the position vectors of two quantum particles
representing the strands and $\vec{r}=\vec{r}_1-\vec{r}_2$ is the mutual position.
We shall extensively use this quantum language.  If there is a gap
between the lowest two eigenvalues of $H$, the
ground state wave-function $\Psi$ determines
the monomer density as $n(\vec{r_1}, \vec{r_2})=\Psi^2(\vec{r}_1,\vec{r}_2)$.
The eigenvalue $E$ is the energy of the quantum pair, related
to the free energy $F$ of the polymer as $E=l\beta^2F/(2N)$.

The surface is described by an infinite potential wall,
\BEA
\label{4}
\Psi(\vec{r}_1,\vec{r}_2)=0, \quad {\rm if}\quad z_1\leq 0, \quad
{\rm or}\quad z_2\leq 0.
\EEA
Both $V(z)$ and $U(r)$ are attractive, $V\le 0,\,U\le 0$, and short-ranged:
$\int_0^\infty\d zV(z)$ and $\int_0^\infty\d rr^2U(r)$ are finite.
When $U=0$, the Hamiltonian $H$ reduces to two uncoupled strands (or
one-dimensional particles), each in the potential $V(z)$.  If $V(z)$ is
shallow enough, no bound (negative energy) state exists, while the
second-order binding transitions corresponds to adsorption of a single
flexible polymer \cite{GKH}. The physical order-parameter for this
transition is $1/\langle z^2\rangle$, which is finite (zero) in the
adsorbed (desorbed) state.  Denote by $\mu$ the dimensionless coupling
constant of $V=\mu\widetilde{V}$ such that (for $U=0$) the 
adsorption threshold is $\mu_c=1$.  Note that the
adsorption of a single strand DNA is a part of the renaturation
via hybridization \cite{bio}, a known method of genetic systematics.

Analogously, switching off both $V(z)$ and the wall, we shall get a
three-dimensional central-symmetric motion in the potential $U(r)$ which
again is not bound if $U$ is shallow. This second-order unbinding
transition with the order parameter $1/\langle r^2\rangle$ corresponds
to thermal denaturation (strand separation) of the double-strand polymer
\cite{bio,PB}.  Writing likewise $U=\lambda \tilde U$, $\lambda$ is the
dimensionless naturation strength. We take the naturation threshold 
in the bulk to be $\lambda_c=1$.  When the wall is included, i.e.
(\ref{4}) is imposed, the strands loose in the adsorbed phase part of
the entropy needed to denaturate.

As seen from Eq.(\ref{1}), in the $(x,y)$ plane $H$ is both
rotationally and translationally symmetric. Thus,
for the low energy physics
the only relevant
variable coming from that plane is
$\rho=\sqrt{(x_1-x_2)^2+(y_1-y_2)^2}$.
The relevant wavefunction is thus
$\psi(z_1,z_2,\rho)$ and the reduced Hamiltonian $H_r$ reads
\BEA
&&H_r=
H_1+H_2
-\rho^{-1}\partial_\rho
\rho\partial_\rho
+U\left(\sqrt{\rho^2+z^2}\right),\\
&&H_{1,2}\equiv-\half\,\partial^2_{z_{1,2}}
+V(z_{1,2}),\qquad z=z_1-z_2.
\EEA
There is no hope for an exact solution of this model, since it
belongs to the class of three-body problems (the role of the third
particle is being played by the surface; see also below).
We shall estimate the lowest energy $E_0$ of $H_r$ via a
variational principle: $E_0\leq \langle \chi|H|\chi\rangle$, where
$\chi$ is a normalized trial function.
Assume that $V(z)$ (without $U$) has a lowest (negative) energy state
with normalized wave function $\phi(z)$:
$H_{1,2}\phi(z_{1,2})=E_V\phi(z_{1,2})$,  with $\phi(z_{1,2}\leq 0)=0$.
We then make the Ansatz
\BEA
\label{ppt}
\chi(\rho,z_1,z_2)=\omega(z_1,z_2)\,\xi(\rho),\quad
\int_0^\infty \d \rho\rho \xi^2(\rho)=1.
\EEA
where $\omega(z_1,z_2)=\phi(z_1)\phi(z_2)$
and where $\xi(\rho)$ is normalized.
Varying $\langle \chi|H-E|\chi\rangle$ w.r.t. $\xi(\rho)$, we
get an effective {\it two-dimensional} problem:
$-\rho^{-1}\partial_\rho(\rho\,\partial_\rho\, \xi)
+U_{\rm eff}(\rho)\xi
=(E-2E_V)\xi$,
where by definition $E\geq E_0$, and
where $U_{\rm e}(\rho)$ is an effective potential:
\BEA
\label{kokand}
U_{\rm e}(\rho)\equiv\int_0^\infty
U(\,\sqrt{\rho^2+(z_1-z_2)^2}\,)
{\prod}_{\k=1}^2\d z_\k \phi^2(z_\k).
\EEA
Like $U$, $U_{\rm e}$ is attractive and short-ranged.  It is
well known that any (no matter how weak) attractive potential in two
dimensions creates a bound state,
though the localization length of this state 
is exponentially large for small energies \cite{LL}.
Thus there is a
normalizable function $\xi(\rho)$ such that $E_0<E<2E_{V}<0$, i.e.,
there is an overall bound (double-strand and localized next to the
surface) state provided $V(z)$ creates a bound state.  In
the present model strong enough surface potential prevents
denaturation (melting) of the double-strand. Upon taking into account
the neglected features (helicity, self-avoidance or stacking),
the denaturation transition temperature in the
adsorbed phase is enhanced but (presumably) finite.

Next to the threshold $\mu_c=1$ of $V(z)$, $2E_V\equiv -\ka^2$ is small
and the wave-function is almost flat \cite{LL}:
$\phi(z)\simeq \sqrt{2\ka}\,e^{-\ka z}$.
Thus Eq.~(\ref{kokand})  implies that
$U_{\rm e}={\cal O}(\ka)$, so  naively it predicts absence of binding
at the threshold $\ka=0$ of $V(z)$.
But let us study the region where $V(z)$ alone does not create any
bound state, by employing the trial function (\ref{ppt}), where
now $\xi(\rho)$ is a given normalized function 
(its form is irrelevant for us), while $\omega(z_1,z_2)$
is found from the variational equation:
\BEA
\label{oman}
&&\left[\,H_1+H_2
+U_{\rm eff}(|z_1-z_2|)
-E\,\right]\omega(z_1,z_2)
=0,\\
&&\label{katar}
U_{\rm eff}(z)
\equiv\int_0^\infty\d \rho\,\rho\,
\xi^2(\rho)\,U(\,\sqrt{\rho^2+z^2}\,),
\EEA
Equations~(\ref{oman}) and (\ref{katar}) describe two one-dimensional particles
interacting via {\it short-range} potential $U_{\rm eff}
=\lambda\widetilde{U}_{\rm eff}$.
We shall show that Eq.~(\ref{oman}) predicts an overall binding|that
is it predicts $E<0$ and a localized normalizable wave-function
$\omega(z_1,z_2)$|at the threshold $\ka=0$ of the potential $V(z)$.  For a
large $\lambda$ this is expected, since two strongly coupled particles
roughly behave as one with double mass in a double potential, which decreases
the threshold value of $V(z)$ by a factor $4$, i.e., $\mu_c(\infty)=\frac{1}{4}$.
Next we focus on small $\lambda$.
Though the energy $2E_V$ is nearly zero, one can apply an ordinary
perturbation expansion in $\lambda$, since the suitable matrix elements
of $U_{\rm eff}$ appear to be small as well, grace to the small
prefactor $\ka$ in the zero-order wave-functions
$|00\rangle\equiv\phi(z_1)\phi(z_2)$. Noting that $\langle 00|U_{\rm
  eff}|00\rangle={\cal O}(\ka)$ for $\ka\to 0$ and thus negligible,
we get at the second order \cite{LL}
\BEA
\label{gamow}
E=
-\int_0^\infty\d K\,
\frac{|\langle 00|U_{\rm eff}|K\rangle  |^2}{\varepsilon_K+\ka^2},
\EEA
where the integration over the collective variable $K$ involves
all excited wave-functions of
the unperturbed two-particle system with wave-vector $K$ and energy
$\varepsilon_K$.
There are three orthogonal families of these states:
$|0k\rangle=\phi(z_1)\wphi_k(kz_2)$, $|k0\rangle=\wphi_k(kz_1)\phi(z_2)$,
$|k_1k_2\rangle=\wphi_{k_1}(k_1z_1)\wphi_{k_2}(k_2z_2)$, where $\wphi_k(kz)$ are the
corresponding single-particle excited (continuous spectrum)
wave-function of the potential $V(z)$ with the wave-number $k$.
For $\ka\to 0$ we get after some steps the finite result
\BEA
\label{jk1}
&&\frac{E}{\delta ^2}=-128
\int_0^\infty\frac{\d k}{1+k^2}~\left[
\int_0^\infty\d v\,e^{-3v}\wphi_0(kv)
\right]^2 \\
&&-2\int_0^\infty\frac{\d k_1\d k_2}{\half+k_1^2+k_2^2}\left[
\int_0^\infty\d ve^{-v}\wphi_0(k_1 v)
\wphi_0(k_2v)
\right]^2,\nn
\EEA
where $\delta \equiv\int_0^\infty\,\d u\,U_{\rm eff}(u)$.
With $\wphi_0(k
v)=\sqrt{\frac{2}{\pi}}\sin(kv)$ we get from
(\ref{jk1}) the numerical value  $E=-0.45 \ \delta^2$.
Its order of magnitude can be checked from the exact relation:
\BEA
\label{ser}
\langle 00|U_{\rm eff}|\omega\rangle=(E-2E_V)
\langle 00|\omega\rangle,\quad 2E_V\equiv-\kappa^2.
\EEA
Assume that $V(z)$ is close to its threshold (i.e., $\mu\to
1^+$, or $\ka\to 0$) and that a weakly-bound state exists with energy
$E=-p^2$. Recall that $|00\rangle$ is mainly a constant modulo a numerical
factor which cancels from (\ref{ser}).
Since all potentials in (\ref{oman}) are short-range and thus
negligible for
large $z_1$ and $z_2$, we get
$\omega(z_1,z_2)=f(z_1,z_2) e^{-p(z_1+z_2)}$, where $f\to 1$ for
$z_1,z_2\to \infty$. This is put into (\ref{ser}) and for $\ka\to 0$
and small $p$ one gets indeed $E\simeq -\delta^2\sim -\lambda^2$.

For $\mu<1$ and $\lambda<1$ we are in the situation where neither
$V(z_1)+V(z_2)$ (the attractive wall alone) nor $U$ alone can create a
bound state.  We thus conclude from (\ref{jk1}, \ref{ser}) that the
present approach does predict binding for $\mu=1$ and for sufficiently
small $\lambda$.  Since the ground state is supposed to be continuous in
$\lambda$ and $\mu$, the very fact of having a negative energy at
$\mu=1$ and not very large $\lambda$ implies that a naturated, adsorbed
state will also exist for $\mu_{\rm c}(\lambda)<\mu<1$, where neither of
the potentials $V$ and $U$ alone allows binding.  The precise curve
$\mu_{\rm c}(\lambda)$, where the ground state energy is equal to its
value at $V=0$ (adsorption threshold), requires numerical analysis to be
reported elsewhere \cite{Allah}.  We thus have found an example of so
called Borromean binding \cite{Nielsen}, where the involved potentials
do not produce bound states separately, but their cumulative effect does
so.  We see from (\ref{ppt}) that this unusual binding is connected to
correlations between the $z$-components of the particles and
(separately) to correlations between their $x$ and $y$ components (in
$\rho$). 

{\it No-binding conditions.} To complete the phase diagram,
we employ a method suggested in \cite{Fl}.
Let us introduce a third fictive particle with mass $M$ and the radius
vector $\vec{r}_3=(x_3,y_3,z_3)$. This particle will substitute the
surface: $V(z_\k)$ ($\k=1,2$) becomes $V(|z_\k-z_3|)$, and condition (\ref{4})
is put at $z_\k=z_3$ only. The
modified problem with Hamiltonian $H_M$
reduces to the original one when for $M\to\infty$,
$\vec{r}_3$ becomes a fixed vector which can be chosen at the origin. We have
\BEA
H_M=-\frac{\vec{\partial}_3^{\,2}}{2M}+U(r)+ {\sum}_{\k=1}^2[\,
  -\half\,\vec{\partial}_\k^{\,2}+V(|z_\k-z_3|) \,].\nonumber
\EEA
It is invariant when shifting the radius
vectors $\vec{r}_\k$ ($\k=1,2,3$) over any fixed vector, while for a
finite-particle quantum system, a symmetry of $H_M$ implies
the same symmetry for its ground-state wave-function \cite{LL}. Thus,
\BEA
\label{1000}
\Psi=\Psi(\vec{r_1}-\vec{r_2},\,\vec{r_1}-\vec{r_3},\,\vec{r_2}-\vec{r_3}),\quad
{\sum}_{\k=1}^3\vec{\partial}_\k\Psi=0.
\EEA
We now decompose $H_M=H_{0}+H_{12}+H_{13}+H_{23}$, where
\BEA
\label{t2}
&&H_{0}\equiv -\half\left[ a\vec{\partial}_3+b{\sum}_{l=1}^2\vec{\partial}_l
\right]\cdot{\sum}_{\k=1}^3\vec{\partial}_\k,\\
&&H_{13}\equiv
-\frac{c}{2}\left(
\frac{\vec{\partial}_1-x\vec{\partial}_3}{1+x}
\right)^2
+V(|z_1-z_3|),
\\
\label{t4}
&&H_{12}\equiv
-2d\left(
\frac{\vec{\partial}_1-\vec{\partial}_2}{2}
\right)^2+U(|\vec{r}_1-\vec{r}_2|),
\EEA
while $H_{23}=H_{13}(1\to 2)$.
The coefficients $a,b,c$ and $d$ can be read off directly:
$a=mM^{-1}-2x^2/(1+2x)^2$,
$b=d={2x(1+x)}/{(1+2x)^2}$, $c={(1+x)^2}/{(1+2x)^2}$.
Since 
$c$ and $d$ are to be employed as effective masses, for the free
parameter $x$ we consider $x\geq 0$ (the regime $x\leq -1$ is of no help).
Note that $\langle\Psi|H_0|\Psi\rangle=0$
due to (\ref{1000}).
Let us see when $\langle\Psi|H_M|\Psi\rangle>0$, i.e, when
a bound state is absent. Changing the  variables|e.g.,
$\vec{s}_1=(1+x)\vec{r}_1$, $\vec{s}_3=(1+x)\vec{r}_1+\vec{r}_3$
for $H_{13}$|we see that $\langle\Psi|H_{13}|\Psi\rangle$
and $\langle\Psi|H_{12}|\Psi\rangle$
are positive for, respectively,
$\mu\leq c(x)$ and $\lambda\leq 2d(x)$.
There is no binding  under these conditions for any $M$ including
$M\to\infty$ which returns to the original problem.
By varying $x$ from $0$ to $\infty$, we cover the domain
\BEQ \label{DDregion}
 0\le\mu\le\frac{1}{4}(2-\lambda+2\sqrt{1-\lambda}\,) \EEQ
\begin{figure}
\includegraphics[width=0.79\linewidth]{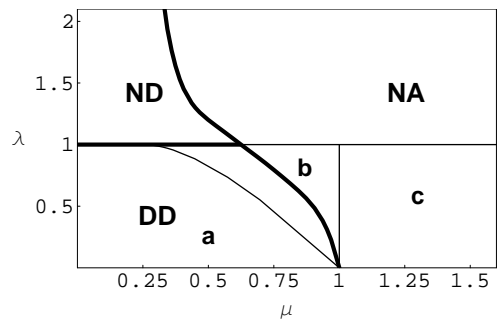}
\hfill
\caption{
Schematic phase diagram for
the inter-strand coupling  $\lambda$ versus 
 the strand-surface coupling $\mu$. The bold lines confine 
three thermodynamical phases.
{\bf ND}: Naturation and desorption.
{\bf NA}: Naturation and adsorption. 
{\bf DD}: Desorption and denaturation.
The critical naturation strength in the
bulk is $\lambda_c=1$, for single strand adsorption it is $\mu_c=1$.
The following subregions are confined by normal lines.
{\bf a}: Domain described by the no-binding condition  (\ref{DDregion}).
{\bf b} (bounded  by the bold {\bf DD}-{\bf NA} line and two straight segments): 
Borromean naturation and adsorption.
{\bf c}: Adsorption and naturation due to overcritical
coupling to the  surface.
}
\label{1f}
\end{figure}
This is the no-binding region $a$ shown on Fig.~\ref{1f}.
At $\lambda=1$ one needs at least $\mu_c=0.25$ to achieve binding,
while at $\mu=1$, $\lambda=0^+$ suffices.  

As shown on Fig.~\ref{1f},
there are three phases in the model: denaturated and desorbed ({\bf
DD}), naturated and adsorbed ({\bf NA}), and naturated and desorbed
({\bf ND}). The denaturated and adsorbed phase is present only for
$U=0$; recall the discussion below (\ref{kokand}).  The exact boundary
between {\bf DD} and {\bf NA} phases has to be convex: if
$(\lambda_1,\mu_1)$ and $(\lambda_2,\mu_2)$ are in {\bf DD}|i.e.,
$H(\lambda_1,\mu_1)\geq 0$ and $H(\lambda_2,\mu_2)\geq 0$|then the
entire line joining these points is in {\bf DD}, because $\lambda$ and
$\mu$ enter $H$ linearly: $H(\epsilon\lambda_1+(1-\epsilon)\lambda_2
,\epsilon\mu_1+(1-\epsilon)\mu_2)\geq 0$, where $1>\epsilon>0$. 

We already mentioned that for large $\lambda$ the adsorption of naturated
DNA will happen at $\mu_c=\frac{1}{4}$ due to doubling of the mass and
potential. For $\lambda$ approaching $\lambda_c=1$ from above, the
absorption is weaker and the entropy will increase so we expect that
the potential $U$ will facilitate adsorption less, leading to the {\bf
ND-NA} curve with $\mu_c>\frac{1}{4}$; see Fig. 1. At $\lambda=1$ it
is expected to meet {\bf DD-NA} at a point, that
separates the three phases {\bf ND}, {\bf NA} and {\bf DD}. 
The {\bf DD-ND} line is implied by the fact that in the 
desorbed phase the polymer is far from the wall.

{\it In conclusion}, we modeled the surface adsorption of a
double-stranded DNA.  There are several situations where the description
of this process via a single chain will not be adequate, and the
two-chain modeling is needed.  (1) The adsorption realized by the same
hydrogen-bonding mechanism as the naturation \cite{review}. Then in the
vicinity of the (temperature or pH induced) melting transition also the
binding to the surface will be small. Thus the motion of separate chains 
will be resolved. (2) The binding to hydrophobic
surfaces (e.g., aldehyde-derivate glass) goes via partial melting which
exposes the hydrophobic core of the helix and leads to the DNA-surface
attraction \cite{review}. Both naturation and adsorption are
simultaneously weakened by increasing the pH \cite{review}.  (3) For homogeneous DNA
at normal conditions (pH$=7$ and NaCl concentration of $0.15$ M) the thermal
melting occurs at temperatures 67 C and 110 C for $A$-$T$ and $C$-$G$
unbinding, respectively \cite{bio,GKH}.  This temperature can be radically
decreased by increasing the pH factor, i.e., by decreasing the
concentration of free protons in the solvent, since the negatively charged
phosphate groups on each strand are not screened any more by protons and
strongly repel each other \cite{bio}. Due to the same reason, for the DNA
adsorption on a positively charged surface \cite{SPM}, the increase of
the pH will increase the electrostatic attraction to the surface. Thus
weakly naturated and weakly adsorbed states may be produced by
controlling the pH and the surface charge. 

Our model describes DNA strands as flexible homopolymers. Many features
of real DNA are put aside in this way: stacking, self-avoidance,
helicity, heterogeneity of the base-sequence, dependence of the
surface-strand interaction on the denaturation degree (in 
the naturated state the bases 
do not participate in the surface-strand
interaction), etc. However, the model can still be useful in clarifying
the mechanisms of DNA adsorption/naturation.  We saw that a
sufficiently strongly attracting surface prevents thermal denaturation
(melting) of the double-strand, because effectively the problem reduces
to binding in two dimensions, where any attractive potential creates a
bound state.  In practice this implies that in the presence of the
surface the melting temperature increases.  It was shown that the
attractive surface and inter-strand coupling together can create bound
states (i.e., double-stranded and adsorbed to the surface), even when
neither of these interactions alone is capable of creating bound states.
This unusual type of binding is related to finely correlated motion of
the strands.  Such Borromean states, where binding is due to the
cumulative effect of several potentials, first appeared in nuclear
physics \cite{Nielsen,Fl}, but their detection in that field is
difficult.  Their observation in polymer physics might be easier, since
the involved scales are mesoscopic.  
Our results led to the schematic phase diagram which, as we argued, may be
checked experimentally.



Thanks to Y. Mamasakhlisov for discussions.
A. E. A. and Zh. S. G. were supported by FOM/NWO
and A. E. A. also by the CRDF grant ARP2-2647-YE-05.
C. K. H. was supported by NSC (Taipei) Grants
NSC 93-2112-M 001-027 \& NSC 94-2119-M-002-001,
 and Academia Sinica (Taipei) Grant
 AS-92-TP-A09.

\end{document}